# What is the boundary condition for the radial wave function of the Schrödinger equation?

Anzor A. Khelashvili[a]

Institute of High Energy Physics, Iv. Javakhishvili Tbilisi State University, University Str. 9, 0109, Tbilisi, Georgia and St. Andrea the First-called Georgian University of Patriarchy of Georgia, Chavchavadze Ave. 53a, 0162, Tbilisi, Georgia

Teimuraz P. Nadareishvili[b]

Institute of High Energy Physics, Iv. Javakhishvili Tbilisi State University, University Str. 9, 0109, Tbilisi, Georgia

**Abstract.** There is much discussion in the mathematical physics literature as well as in quantum mechanics textbooks on spherically symmetric potentials. Nevertheless, there is no consensus about the behavior of the radial function at the origin, particularly for singular potentials. A careful derivation of the radial Schrödinger equation leads to the appearance of a delta function term when the Laplace operator is written in spherical coordinates. As a result, regardless of the behavior of the potential, an additional constraint is imposed on the radial wave function in the form of a vanishing boundary condition at the origin.

**Keywords:** Schrödinger equation, radial equation, boundary condition, singular potentials.

PACS numbers: 03.65.-w, 03.65.Ca, 03.65.Ta

## I. INTRODUCTION

According to the general principles of quantum mechanics, the wave functions must obey certain requirements, such as continuity (more precisely, two-fold differentiability), uniqueness, and square integrability. In many problems knowledge of the behavior of the wave function is needed at points where the potential has a singularity. In this paper we consider spherically symmetric potentials for which separation of variables is performed in spherical coordinates. It is well known that the transformation to spherical coordinates is a singular at the origin. The transformation from Cartesian to spherical coordinates is not unambiguous, because the Jacobian of this transformation is $J = r^2 \sin\theta$ and is singular at $r = 0$ and $\theta = n\pi (n = 0,1,2,...)$. The angular part is unambiguously fixed by the requirements of continuity and uniqueness[1] and gives the spherical harmonics $Y_l^m(\theta,\varphi)$.

We note that though $\vec{r} = 0$ is an ordinary point in the full Schrödinger equation, it is a singular point in the radial equation and thus, knowledge of the behavior at $\vec{r} = 0$ is required. We consider the radial wave function $u(r)$, which is a solution to

$$\frac{d^2 u(r)}{dr^2} - \frac{l(l+1)}{r^2} u(r) + 2m[E - V(r)]u(r) = 0. \qquad (1)$$

Equation (1) includes only the second derivative. It is clear from Eq. (1) that the behavior of $u(r)$ at the origin depends on the behavior of the potential V(r), in particular, whether it is regular or singular. Although a definite answer exists for regular potentials, the situation is unclear for singular potentials.[2] We reconsider the derivation of the radial equation in more detail and show that the existence of the radial equation depends on the behavior of $u(r)$ at the origin.

This article is organized as follows. In Sec. II we consider the consequences of some general principles. We will show that there is no unambiguous answer. In Sec. III we consider the transformation to the radial equation and obtain a additional delta-like term, elimination of which provides a constraint on the radial wave function at the origin. Only after satisfying this constraint does the radial equation take its usual form, which is Eq. (1). This constraint has the form of a boundary condition for the radial wave function at the origin. In Sec. IV we give concluding remarks, and in the Appendix we discuss the appearance of the delta function in the radial equation.

## II. THE BOUNDARY CONDITION AT THE ORIGIN

The question is what is the maximal singularity that the radial function $R(r)$ or $u(r) = rR(r)$ can have at the origin $r = 0$. The complete three-dimensional wave function is

$$\psi(\vec{r}) = R(r) Y_l^m(\theta, \varphi); \quad r > 0; \quad 0 \leq \theta \leq \pi; \quad 0 \leq \varphi \leq 2\pi, \qquad (2)$$

and the equation for the function $R(r)$ is

$$\frac{d^2 R}{dr^2} + \frac{2}{r}\frac{dR}{dr} + 2m[E - V(r)]R - \frac{l(l+1)}{r^2} R = 0. \qquad (3)$$

The traditional change of variables eliminates the first derivative term from Eq. (3) by the substitution

$$R(r) = \frac{u(r)}{r}, \qquad (4)$$

which leads to Eq. (1) for the radial wave function $u(r)$. We will refer to u(r) as the radial wave function.

From the continuity of $R(r)$ at $r = 0$ it follows that $u(0) = 0$, insuring a finite probability at this point.[3] We can weaken this condition by requiring a finite differential probability in the spherical slice $(r, r + dr)$

$$|R|^2 r^2 dr < \infty. \qquad (5)$$

If $R \sim r^s$ at the origin, it follows that $s > -1$, or $u(0) = 0$.

Another generalization is to require a finite total probability inside a sphere of small radius $a$,

$$\int_0^a |R|^2 r^2 dr < \infty. \tag{6}$$

In this case more singular behavior is permissible, namely,

$$\lim_{r \to 0} u(r) \approx \lim_{r \to 0} r^{-1/2+\varepsilon} \to \infty, \tag{7}$$

where $\varepsilon > 0$ is a small positive constant and $\varepsilon \to 0$ at the end of the calculation.

The same behavior follows from the finite behavior of the norm.

$$\int_0^\infty |R(r)|^2 r^2 dr < \infty. \tag{8}$$

We can also use a stronger argument by Pauli,[4] namely, the time independence of the norm. To explore it we follow the procedure in Ref. 5. In quantum mechanics the norm of the wave function is independent of time

$$\frac{d}{dt} \int \psi^* \psi \, dV = 0. \tag{9}$$

By using the time-dependent Schrödinger equation, we transform Eq. (9) to

$$-\frac{i}{\hbar} \int \left[ \psi^*(H\psi) - (H\psi)^* \psi \right] dV = 0. \tag{10}$$

The time independence of the probability means that the Hamiltonian must be a Hermitian operator. By introducing the probability current density

$$\vec{J} = \operatorname{Re}\left[ \psi^* \frac{\hbar}{im} \vec{\nabla} \psi \right], \tag{11}$$

it is easy to show that

$$\operatorname{div} \vec{J} = \frac{i}{\hbar} \left[ \psi^*(H\psi) - (H\psi)^* \psi \right]. \tag{12}$$

The equation for conservation of probability takes the form (after using Gauss' theorem)

$$\frac{d}{dt}\int_V \psi^*\psi \, dV = -\int_V \text{div}\vec{J} \, dV = -\int_S J_N \, dS \tag{13}$$

where $J_N$ is the normal component of the current relative to the surface.

If we assume that at $r = 0$ the Hamiltonian has a singular point, Gauss' theorem in Eq. (13) is not applicable. We must exclude this point from the integration volume and surround it by a small sphere of radius $a$. In this case the surface integral is divided into a surface at infinity that encloses the total volume, and the surface of a sphere of radius $a$:

$$\lim_{a \to 0} a^2 \int J_a \, d\Omega + \int_\infty J_N \, ds = 0. \tag{14}$$

In the first integral in Eq. (14) we have expressed the surface element of the sphere as $ds = a^2 d\Omega$, where $d\Omega$ is an element of solid angle. Because the wave function must vanish at infinity, the second term goes to zero. If we substitute $J_a = \frac{i\hbar}{2m}\left(\psi \frac{\partial \psi^*}{\partial r} - \psi^* \frac{\partial \psi}{\partial r}\right)_{r=a}$ and assume $\psi = \frac{\tilde{u}}{r^s}$, where $\tilde{u}$ is regular at $r \to 0$, we obtain

$$\lim_{a \to 0} \frac{a^2}{a^{2s}} \int \left(\tilde{u} \frac{\partial \tilde{u}^*}{\partial r} - \tilde{u}^* \frac{\partial \tilde{u}}{\partial r}\right)_{r=a} d\Omega = 0 \tag{15}$$

Equation (15) is satisfied if $s < 1$. It follows that $R(r)$ does not diverge more quickly than $1/r^s$, with $s < 1$, which means that $\lim_{r \to 0} u(r) \approx \lim_{r \to 0} r^{-s+1} \to 0$.

We see that the different arguments lead to different conclusions for the wave function behavior at the origin. A finite norm allows for divergent behavior of $u(r)$ at the origin, but the time independence of the norm gives vanishing behavior.

Does the boundary behavior at the origin have some physical meaning? To discuss this question we start from the Eq. (1) and consider the well known example of a regular potential

$$\lim_{r \to 0} r^2 V(r) = 0. \tag{16}$$

After we substitute $u \sim r^s$ near the origin, it follows from the characteristic equation that $s(s-1) = l(l+1)$, which gives two solutions: $u \underset{r \to 0}{\sim} c_1 r^{l+1} + c_2 r^{-l}$. For nonzero $l$ the second term is not locally square integrable and is usually ignored. Many authors discuss how to deal with the solution for $l = 0$,[6,7] which is square integrable at the origin. Messiah[7] writes: "The foregoing argument does not apply when $l = 0$. But in that case, the corresponding wave function $\psi_0$ ($R_0$ in our notation) does not satisfy the Schrödinger equation [condition (a)]. In fact, $\psi_0$ behaves as $(1/r)$ at the origin, and since $\nabla^2 (1/r) = -4\pi \delta(\vec{r})$,

$$(H - E)\psi_0 = \frac{2\pi \hbar^2}{m} \delta(\vec{r}). \tag{17}$$

One must therefore keep only the so-called "regular" solutions, that is, the solutions satisfying the condition $[u(0) = 0]$. With such a solution we can be sure that the function $\psi_l^m$ is a solution of the Schrödinger equation everywhere, including the origin".

However this consideration corresponds only to a regular potential. The analysis changes drastically when the potential is singular. Consider the following singular potential

$$\lim_{r \to 0} r^2 V(r) = -V_0 = \text{constant}, \tag{18}$$

where $V_0 > 0$ corresponds to attraction and $V_0 < 0$ corresponds to repulsion.

For this potential the equation for the exponent a takes the form

$s(s-1) = l(l+1) - 2mV_0$, which has two solutions: $s = \frac{1}{2} \pm \sqrt{\left(l + \frac{1}{2}\right)^2 - 2mV_0}$.

Therefore

$$u \underset{r \to 0}{\sim} c_1 r^{\frac{1}{2}+P} + c_2 r^{\frac{1}{2}-P}; \quad P = \sqrt{\left(l + \frac{1}{2}\right)^2 - 2mV_0}. \tag{19}$$

Both solutions are square integrable near the origin as long as $0 \leq P < 1$. This condition is studied in connection with the self-adjoint extension of the radial Hamiltonian.[8,9] It corresponds to the condition in Eq. (6). For the condition in Eq. (5), $P$ is restricted to $0 \leq P < 1/2$. The difference in the upper bound is essential. The radial equation takes the form

$$u''(r) - \frac{P^2 - 1/4}{r^2} u(r) + 2mEu(r) = 0. \tag{20}$$

Depending on whether $P$ is greater then 1/2 or not, the sign in front of the fraction in Eq. (20) changes, and we can derive the results for an attractive potential using the case of a repulsive potential and vice versa, whereas the condition in Eq. (5) forbids this undesirable case with $1/2 \leq P < 1$.

### III. WHEN IS THE RADIAL EQUATION VALID?

It seems that the choice $u(0) = 0$ is preferable. But this condition does not follow directly from Eq. (1). Therefore we reconsider the derivation of Eq. (1) in more detail. We return to the rigorous derivation of the radial equation for $u(r)$. After substitution of Eq. (3) into Eq. (2) we obtain

$$\frac{1}{r}\left(\frac{d^2}{dr^2} + \frac{2}{r}\frac{d}{dr}\right)u(r) + u(r)\left(\frac{d^2}{dr^2} + \frac{2}{r}\frac{d}{dr}\right)\left(\frac{1}{r}\right) + 2\frac{du}{dr}\frac{d}{dr}\left(\frac{1}{r}\right) - \left[\frac{l(l+1)}{r^2} - 2m(E - V(r))\right]\frac{u}{r} = 0 \tag{21}$$

We write the radial equation in the form (21) to show the action of the radial part of the Laplacian explicitly. The first derivatives of $u(r)$ cancel, and we are left with

$$\frac{1}{r}\left(\frac{d^2u}{dr^2}\right)+u\left(\frac{d^2}{dr^2}+\frac{2}{r}\frac{d}{dr}\right)\left(\frac{1}{r}\right)-\frac{l(l+1)}{r^2}\frac{u}{r}+2m(E-V(r))\frac{u}{r}=0. \quad (22)$$

As we do the derivatives in the second term naively, we obtain zero, when $r\neq 0$. If we take into account that

$$\frac{d^2}{dr^2}+\frac{2}{r}\frac{d}{dr}=\frac{1}{r^2}\frac{d}{dr}\left(r^2\frac{d}{dr}\right)\equiv\nabla^2_r \quad (23)$$

is the radial part of the Laplacian, we conclude that[10]

$$\nabla^2_r\left(\frac{1}{r}\right)=\nabla^2\left(\frac{1}{r}\right)=-4\pi\delta^{(3)}(\vec{r}), \quad (24)$$

and thus Eq. (22) becomes

$$\frac{1}{r}\left[-\frac{d^2u(r)}{dr^2}+\frac{l(l+1)}{r^2}u(r)\right]+4\pi\delta^{(3)}(\vec{r})u(r)-2m[E-V(r)]\frac{u(r)}{r}=0 \quad (25)$$

It includes an extra three-dimensional delta-function term, which is evident from Eq. (24), and discussed in the Appendix. Its presence in the radial equation has no physical meaning and thus it must be eliminated. Note that if $r\neq 0$, this extra term vanishes due to the nature of the delta function. If $r\neq 0$ and we multiply Eq. (24) by $r$, we obtain the ordinary radial equation (1).

If $r=0$, multiplication by $r$ is not permissible and the extra term remains in Eq. (25). Therefore we have to investigate this term separately and find a way to discard it.

The effect of the three-dimensional delta function is determined by integrating over $d^3r=r^2dr\sin\theta d\theta d\varphi$. It is evident that[10]

$$\delta^{(3)}(\vec{r}) = \frac{1}{|J|}\delta(r)\delta(\theta)\delta(\varphi) \qquad (26)$$

where $J = r^2 \sin\theta$ is the Jacobian. Thus, the extra term effectively becomes

$$u(r)\delta^{(3)}(\vec{r})d^3\vec{r} \to u(r)\delta(r)dr. \qquad (27)$$

Its appearance as a point-like source at r = 0 is not physical. The only reasonable way to remove this term without modifying the Laplace operator or including a compensating delta function term in the potential $V(r)$, is require that

$$u(0) = 0. \qquad (28)$$

Multiplication of Eq. (25) by $r$ and elimination of the delta function due to the property $r\delta(r) = 0$ is not acceptable, because it is equivalent to multiplication of this term by zero. Therefore we conclude that the radial equation (1) for $u(r)$ is compatible with the full Schrödinger equation (2) if and only if the condition $u(0) = 0$ is satisfied.

Equation (1) supplemented by the condition (28) is equivalent to Eq. (2). It satisfies the Dirac requirement[11] that the solutions of the radial equation must be compatible with the full Schrödinger equation. It is remarkable that the supplementary condition (28) has the form of a boundary condition at the origin. All of these statements can be easily verified by explicit integration of Eq. (9) over a small sphere with radius $a$ approaching to zero at the end of the calculations.

We have already seen that there is some ambiguity in the formulation of the boundary condition for the radial wave function from the general principles of quantum mechanics. Therefore various boundary conditions have been considered, especially for singular potentials. We have shown that the radial equation is valid only together with condition (28), independently of the potential, whether it is regular or singular.

Usually boundary conditions are derived from the radial equation for a given potential. But our result means that the radial equation (1) by itself follows from the total Schrodinger equation if and only if the constraint (28) is satisfied. It is curious that this fact (appearance of delta functions while reducing the Schrödinger equation) has apparently gone unnoticed.

Previous papers that have explored this boundary condition are obviously correct.[12,13] In contrast, papers without this boundary condition are doubtful, because the Eq. (1) is valid only if Eq. (28) is satisfied. Most textbooks consider only regular potentials with this boundary condition and therefore their results are correct. The only exception is the $l = 0$ state, which has been discussed by Messiah.[7] We proved his assumption, because the second solution must be ignored for any $l$, including $l = 0$.

More far-reaching consequences follow for singular potentials. Many authors[8,9,14] neglect the boundary condition entirely and satisfy only square integrability. But in this treatment some of parameters of wave functions go out of allowed regions and a self-adjoint extension procedure can yield unphysical results. A corresponding example was mentioned after Eq. (20) for $1/2 \leq P < 1$, where a repulsive potential gives a bound state after a self-adjoint extension.[9] Other examples of singular potentials are considered in Ref. 15.

## IV.   CONCLUSIONS

We have shown that a rigorous reduction of the Laplace operator in spherical coordinates leads to a previously unnoticed delta function term. Careful investigation of this term gives a constraint on the behavior of the radial wave function at the origin in the form of a boundary condition, $u(0)=0$. A unique boundary condition follows for both regular and singular potentials. Only the nature of the approach to zero depends on the behavior of the potential at the origin.

Since at least the work of Case,[12] it has been known the importance of notions of limit-point, limit cycle and self-adjoint extension procedure for the radial Schrodinger equation and it's Hamiltonian.[16,17] It provides the correct way to understand the boundary conditions at the origin for Eq. (1). There is nothing wrong with such a treatment, which yields the condition $u(0)=0$ by applying powerful mathematics.[16,17,18] But as we have shown, the radial equation (1) has nothing in common with physics without the condition (28). A self-adjoint extension, used in many papers that do not satisfy this condition, has only mathematical importance.

Similar issues arise in classical electrodynamics,[19] where the extra delta function appears in calculations of dipole electric and magnetic fields, but cancels without any physical consequences. The situation in quantum mechanics differs because the extra delta term necessitates the restriction of the radial wave function. The same issue holds for the radial reduction of the Klein-Gordon equation, because in three dimensions it has the form

$$(-\nabla^2 + m^2)\psi(\vec{r}) = [E - V(r)]^2 \psi(\vec{r}), \qquad (29)$$

and the reduction of variables in spherical coordinates proceeds in the same way as for the Schrödinger equation.

**ACKNOWLEDGEMENTS**

We wish to thank Profs. John Chkareuli, Sasha Kvinikhidze, and Parmen Margvelashvili for valuable discussions. We are also indebted to Prof. Boris Arbuzov, Drs. Irakli Machabeli, and Shota Vashakidze for reading the manuscript.

**APPENDIX : HOW THE DELTA FUNCTION APPEARS**

Following Ref. 10 we show how the delta function appears in the radial equation. Consider the following derivative:

$$\left(\frac{d^2}{dr^2}+\frac{2}{r}\frac{d}{dr}\right)\left(\frac{1}{r}\right). \tag{A1}$$

A naive calculation would yield zero. But the separate terms in this expression are highly singular, and therefore we must regularize them. We choose the following regularization near the origin

$$\frac{1}{r} \to \lim_{a \to 0} \frac{1}{\sqrt{r^2+a^2}}. \tag{A2}$$

Equations (A1) and (A2) lead to

$$\left(\frac{d^2}{dr^2}+\frac{2}{r}\frac{d}{dr}\right)\left(\frac{1}{r^2+a^2}\right)^{1/2} = -\frac{3a^2}{\left(r^2+a^2\right)^{5/2}}. \tag{A3}$$

The right-hand side of Eq. (A3) is well behaved everywhere for $a \ne 0$, but as $a \to 0$ it becomes infinite at $r=0$ and vanishes for $r \ne 0$. To make the connection to a delta function we integrate the right-hand side of Eq. (A3) by $d^3\vec{r} = r^2 dr d\Omega$, which gives

$$-4\pi \int \frac{3a^2}{\left(r^2+a^2\right)^{5/2}} r^2 dr. \tag{A4}$$

We divide the volume of integration into two parts: a sphere of radius $R$ with center at the origin and region outside the sphere. Because $a << R$ and approaches zero, the integral from the exterior of the sphere vanishes as $a^2$ as $a \to 0$. We thus need to consider only the contribution from inside the sphere. We can neglect $r^2$ in the denominator, because the integrand varies very slowly with $r$. After this neglect the integral is equal to

$$\frac{3a^2}{(a^2)^{5/2}} \frac{a^3}{3} = \frac{a^5}{a^5} = 1. \tag{A5}$$

Thus we have all the properties of the 3-dimensional delta function, and we confirm Eq.(24).